\documentclass[aps,pre,twocolumn,superscriptaddress,showpacs]{revtex4-1}

\usepackage{verbatim}
\usepackage{graphicx}
\usepackage{amssymb,mathtools}
\usepackage{float}
\usepackage{hyperref}
\hypersetup{
colorlinks,
citecolor=blue,
filecolor=blue,
linkcolor=blue,
urlcolor=blue,
pdftitle=title,
pdfauthor=author,
bookmarks,}

\DeclareMathOperator{\sign}{sign}

\begin{document}

\title{Slow axial drift in three-dimensional granular tumbler flow}

\author{Zafir Zaman}
\affiliation{Department of Chemical \& Biological Engineering, Northwestern University, Evanston, Illinois 60208, USA}

\author{Umberto D'Ortona}
\affiliation{Laboratoire M2P2, UMR 7340 CNRS / Aix-Marseille Universit\'e, 13451 Marseille Cedex 20, France}

\author{Paul B. Umbanhowar}
\affiliation{Department of Mechanical Engineering, Northwestern University, Evanston, Illinois 60208, USA}

\author{Julio M. Ottino}
\affiliation{Department of Chemical \& Biological Engineering, Northwestern University, Evanston, Illinois 60208, USA}
\affiliation{Department of Mechanical Engineering, Northwestern University, Evanston, Illinois 60208, USA}
\affiliation{The Northwestern Institute on Complex Systems (NICO), Northwestern University, Evanston, Illinois 60208, USA}

\author{Richard M. Lueptow}
\email{r-lueptow@northwestern.edu}
\affiliation{Department of Mechanical Engineering, Northwestern University, Evanston, Illinois 60208, USA}
\affiliation{The Northwestern Institute on Complex Systems (NICO), Northwestern University, Evanston, Illinois 60208, USA}

\date{\today}

\begin{abstract}
Models of monodisperse particle flow in partially filled three-dimensional tumblers often assume that flow
along the axis of rotation is negligible. We test this assumption, for spherical and double cone tumblers, using
experiments and discrete element method simulations. Cross sections through the particle bed of a spherical
tumbler show that, after a few rotations, a colored band of particles initially perpendicular to the axis of rotation
deforms: particles near the surface drift toward the pole, while particles deeper in the flowing layer drift toward
the equator. Tracking of mm-sized surface particles in tumblers with diameters of 8–14 cm shows particle axial
displacements of one to two particle diameters, corresponding to axial drift that is 1–3\% of the tumbler diameter,
per pass through the flowing layer. The surface axial drift in both double cone and spherical tumblers is zero at
the equator, increases moving away from the equator, and then decreases near the poles. Comparing results for
the two tumbler geometries shows that wall slope causes axial drift, while drift speed increases with equatorial
diameter. The dependence of axial drift on axial position for each tumbler geometry is similar when both are
normalized by their respective maximum values
\end{abstract}


\maketitle

\section{Introduction}

Flow of granular media can be difficult to predict and challenging to model because of the inherent complexity of the collective motion of large numbers of particles. Yet, granular flows can sometimes be modeled using a continuum approximation, for example, as in the case of flow in a rotating tumbler~\cite{MeierAdvPhys2007} and flow down an inclined plane or heap~\cite{GDRMidi2004}. The usual assumption in these geometries is that spanwise particle motion is primarily diffusive and averages to zero. However, there are situations where spanwise flow occurs. For example, in a partially-filled, long cylindrical tumbler rotating about its axis, endwall friction causes pathlines to curve near the endwalls~\cite{ManevalGranMatt2005,PengfeiPRESubsurface2008}.
In the case of bidisperse particles, this out-of-plane flow can drive
axial segregation and initiate axial band formation \cite{ZikPRL1994,HillPRE1995,HillPRL1997,ChooPRL1997,AransonPRL1999,ChenPRL2010,PengfeiNJOP2011}.

\begin{figure}
\begin{center}
\includegraphics[width=1\linewidth]{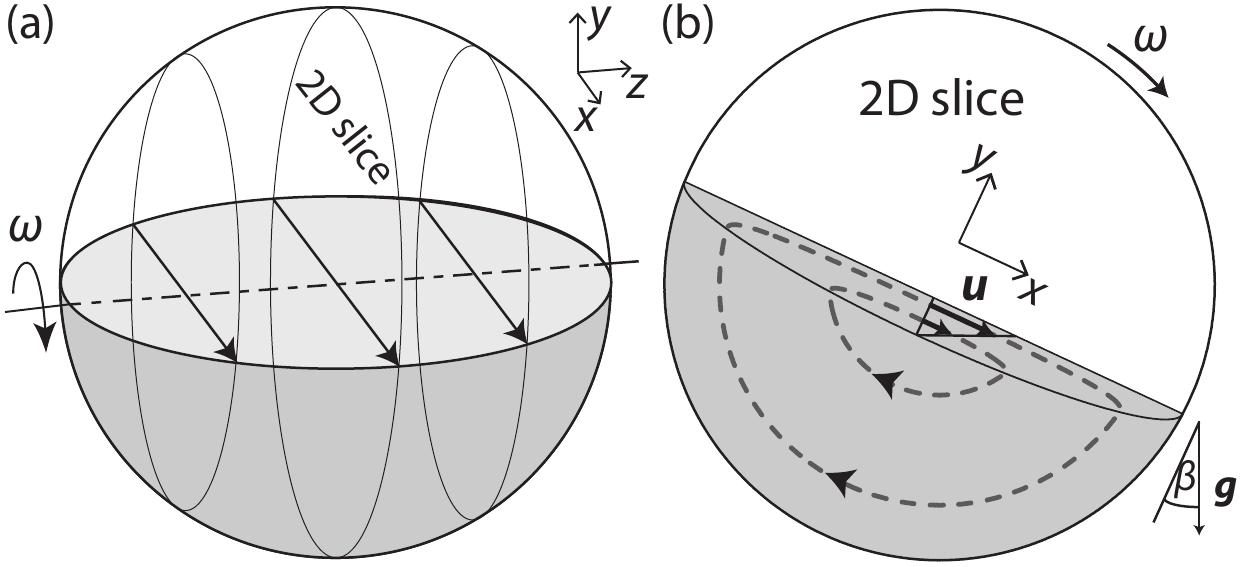}
\caption{(a) Sketch of spherical tumbler flow modeled as a composite of two-dimensional slices. (b) Sketch of a two-dimensional slice model of circular tumbler flow in the continuous flow regime with a thin flowing layer near the surface and solid body rotation for material in the fixed bed. $\omega$ and $\mathbf{u}$ denote the rotation speed and flowing layer velocity, respectively, and $\beta$ is the dynamic angle of repose of the free surface with respect to horizontal.}
\label{figmodel}
\end{center}
\end{figure}

\begin{figure*}[t!]
\includegraphics[width=\textwidth]{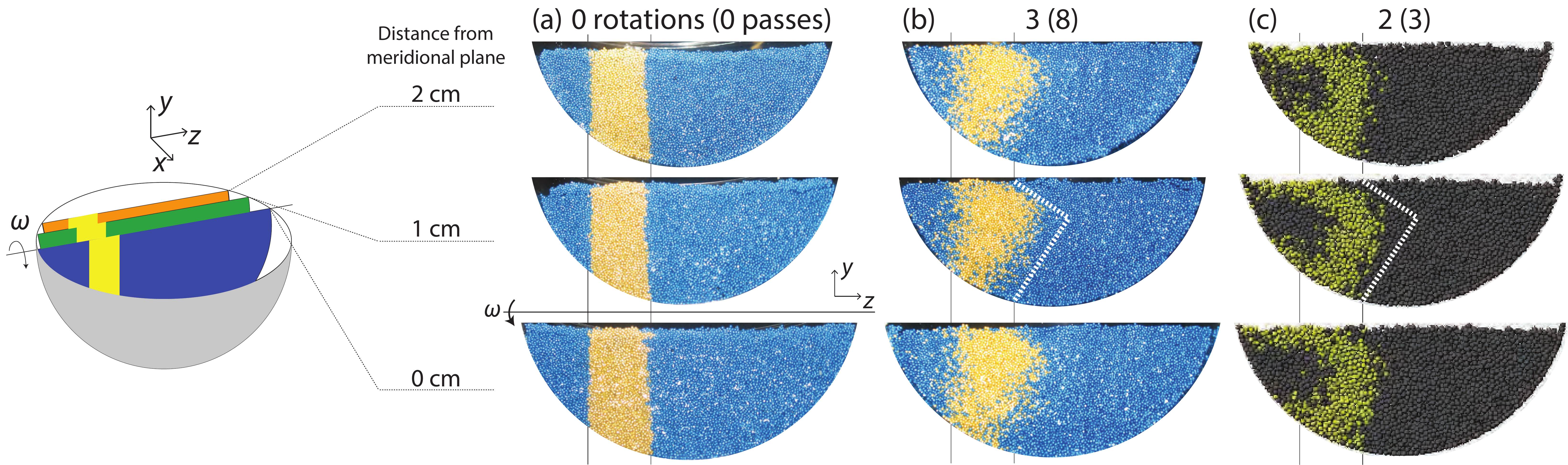}
\caption{(Color online) Axial drift of particles demonstrated by band deformation in $yz-$plane cross sections in a spherical tumbler ($D=14$\,cm) with fill fraction $f=30\%$ from experiment with $d=1$\,mm, (a) zero rotations and (b) three rotations (eight fixed bed passes), and from (c) DEM simulation with $d=2$\,mm after two rotations (three fixed bed passes). The colored band was initially located between $z=-4.3$\,cm and $z=-1.8$\,cm (measured axially from the tumbler equator) and spanned the tumbler in the $x$ direction. Drift occurs throughout the tumbler in the $x$ direction. White dotted lines on the 1-cm cross section highlight the similarity between band deformation in (b) experiment and (c) simulation, though drift per pass is greater in simulation.}
\label{figcrosssection}
\end{figure*}

Here we consider flow in partially-filled three-dimensional (3D) tumblers, such as spherical tumblers [Fig.~\ref{figmodel}(a)] \cite{MeierAdvPhys2007,SturmanMeierJFM2008,GabeEPL2010,JuarezChristovCES2012}, rotating with angular velocity $\omega$ about a horizontal axis $z$ that intersects the tumbler at its ``poles.'' Our interest derives from the desire to extend the understanding of granular flow in quasi-two-dimensional (2D) cases~\cite{CantelaubeEPL1995,ClementEPL1995,HillIntlBif1999,KhakharJFM2001,HillPRL2003,JainJFM2004,GDRMidi2004} to fully 3D flows. We consider the situation where the free surface is essentially flat and continuously flowing~\cite{MeierAdvPhys2007}. In this regime, the surface of the flowing layer maintains a dynamic angle of repose with respect to horizontal $\beta$ which depends on the frictional properties and diameter $d$ of the particles, and the rotational speed of the tumbler $\omega$~\cite{TaberletPRL2003,CourrechEPL2003,DopplerJFM2007,PignatelPRE2012}. A simple model of 3D flow in a spherical tumbler~\cite{MeierAdvPhys2007,GabeEPL2010,JuarezChristovCES2012} assumes that flow in each plane perpendicular to the axis of rotation is essentially that of a two-dimensional circular slice of the appropriate diameter, as shown in Fig.~\ref{figmodel}(a). In this reduced quasi-2D geometry, particles enter the upstream end of the thin, rapidly flowing layer from the fixed bed (the region of particles in solid body rotation), flow downslope, and return to the fixed bed following the idealized streamlines shown in Fig.~\ref{figmodel}(b).  The streamwise velocity profile in the flowing layer decreases approximately linearly with depth~\cite{KhakharPhysFluids1997,JainPhysFluids2002,OttinoKhakharAIChE2002}, while the portion of the `solid body' region nearest the flowing layer exhibits a much slower creeping motion~\cite{KomatsuPRL2001,SociePRE2005,ArndtPRE2006}.

Despite the attractiveness of its simplicity, there are indications that the 2D flow assumption is imperfect. Simulations of monodisperse flow in a partially-filled spherical tumbler~\cite{PengfeiPRLBandInv2009} indicate a slight out-of-$xy$-plane curvature in the trajectories of surface particles. Asymmetries between the upstream and downstream portions of the curved trajectories manifest as axial drift. Unlike a partially-filled cylindrical tumbler~\cite{SantomasoCES2004,PohlmanPRE2009}, the trajectory curvature cannot be directly attributed to frictional endwall effects. Rather, the curved particle trajectories appear to be related to the relative curvature of the tumbler walls to the surface of spherical particles as the paths of large particles curve more than the paths of small particles for the same tumbler diameter.

In this paper, axial drift is examined in experiment and simulation. Experimentally, particles are tracked on the surface and along the tumbler wall to measure the axial drift. Cross sections of the tumbler are also imaged to visualize the axial motion of colored particles within the bed. Discrete element method (DEM) simulations are performed to obtain particle trajectories and velocities to understand the axial drift in more detail.  Similar axial drift occurs in both experiment and simulation over a range of tumbler and particle parameters, confirming the robust nature of the phenomenon.

\begin{figure}[!h]
\begin{center}
\includegraphics[width=1\linewidth]{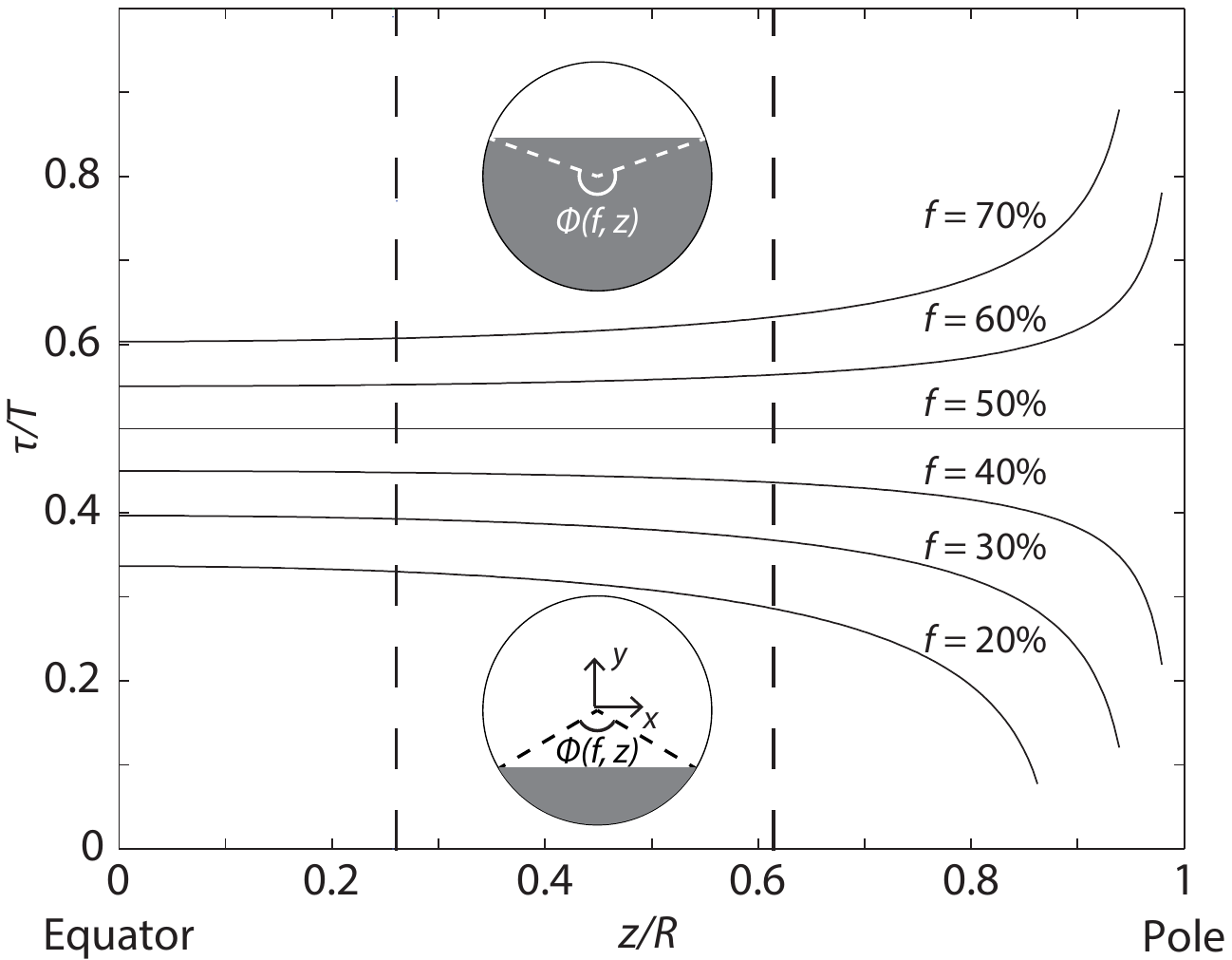}
\caption{Calculated axial variation of normalized solid body rotation time, $\tau/T$, vs.~normalized axial position, $z/R$, for $20\% \leq f \leq 70\%$ in a spherical tumbler. Curves extend from the equator to the maximum axial position of the free surface. The region between vertical dashed lines represents the initial position of the tracer band ($0.257 < z/R < 0.614$) in experiments and simulations.}
\label{figtime}
\end{center}
\end{figure}

\begin{figure*}[!t]
\begin{center}
\includegraphics[width=\textwidth]{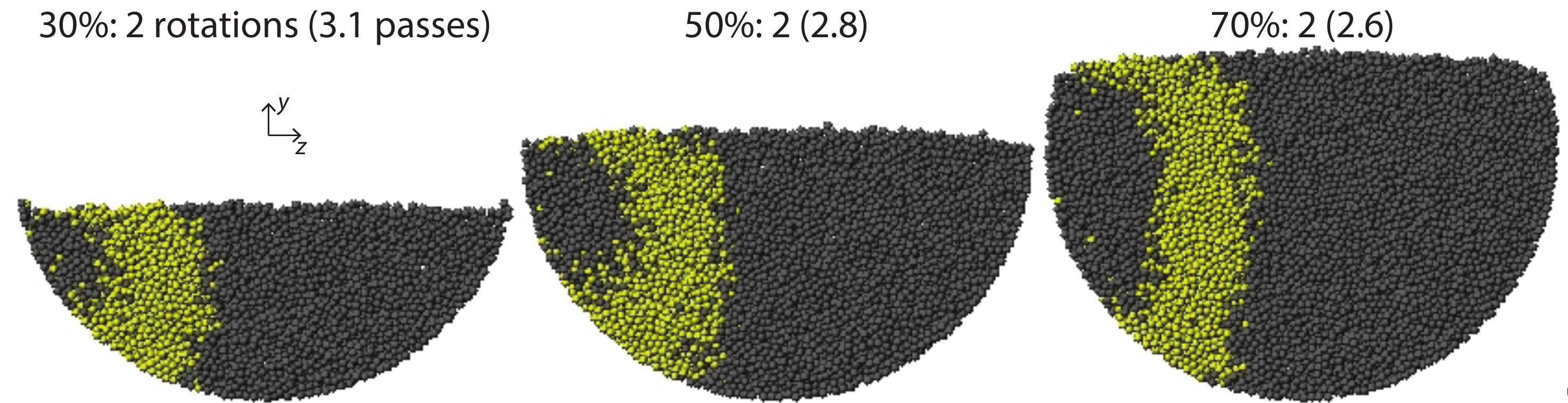}
\caption{(Color online) Band deformation in spherical tumblers is robust to variation in fill fraction as shown in $yz$-plane cross sections containing the axis of rotation from simulation. Fill fraction, rotation number and approximate number of passes through the flowing layer (in parenthesis) are indicated above each image. $d=2$\,mm, $D=14$\,cm, and $\omega=30$\,rpm.}
\label{figfillfrac}
\end{center}
\end{figure*}

\section{Band Deformation}
\label{banddef}
The tumblers in these experiments were clear acrylic spheres rotated at constant angular velocity about a horizontal $z$ axis by a motor. In the band drift experiments, a $D=14$-cm diameter tumbler was filled to a fill fraction (by volume) varying between $20\%$ and $50\%$ with $d=1.02\pm0.08$\,mm diameter soda-lime glass beads (SiLiglit Deco Beads, Sigmund Lindner GmbH, Germany) and rotated at $\omega = 5$ rpm. A tracking band [Fig.~\ref{figcrosssection}(a)] was formed by filling the space between two thin partitions (at $z=-4.3$ cm and $z=-1.8$ cm) with light colored beads and filling the exterior with dark colored beads. To reduce electro-static interactions between particles and the wall, the inside of the tumbler was either wiped with an antistatic wipe (Staticide, ACL Inc., Chicago, IL) or treated with an antistatic spray (SP 610, Sprayon Inc., Cleveland, OH) prior to filling. To reduce inter-particle static charging when the relative humidity was below 50\%, a small amount of deionized water (typically 2-5\,$\mu$L) was allowed to evaporate in the sealed tumbler to increase the relative humidity above 50\%.

To view cross sections on $yz$ planes perpendicular to the flow direction in experiment after tumbling, the grain bed was immobilized by pouring hot gelatin into the tumbler. Once the gelatin cooled to room temperature, the tumbler was placed in a freezer to set the gelatin. The tumbler was then cut in half along the meridional plane and photographed. Additional parallel planes behind the initial cut were exposed by carefully removing particles with a scraper. Two sets of images were obtained for each experiment from the halves created by the initial cut.

Before presenting the experimental results, we note that the time between surface particle passes through the flowing layer $\tau$ is always less than the tumbler rotation period $T$ for $f<100\%$ and, as a consequence, the number of particle passes (cycles) is always greater than the number of tumbler rotations. This occurs because the bed surface in any quasi-2D slice perpendicular to the rotation axis subtends an angle, $\phi < 2\pi,$ measured with respect to the rotation axis (Fig.~\ref{figtime} inset). Ignoring the relatively much shorter time spent in the flowing layer (see Appendix~\ref{restime}), $\tau/T = \phi/2\pi$.  For all $f\ne50\%,$ $\phi,$ and thus $\tau/T,$ varies with axial position since the surface of the particle bed does not extend to the poles. Figure~\ref{figtime} shows the predicted value of $\tau/T$ as a function of normalized axial position $z/R,$ where $R=D/2$ is the tumbler radius. For fixed $z/R,$ $\tau/T$ increases with $f,$ while for fixed $f,$ $\tau/T$ decreases (increases) with $z/R$ for $f>50\%$ ($f<50\%$). At $f=50\%,$ $\tau/T$ $(=0.5)$ is independent of $z/R.$

\begin{table}[h!]
\caption{\label{simtable} DEM simulation parameters}
\begin{ruledtabular}
\begin{tabular}{ll}
Tumbler diameter $D$ (cm) & 10, 14\\
Particle diameter $d$ (mm) & 2\\
Particle density $\rho$ (kg\,m$^{-3}$) & 1308~\cite{DrakeShreveJRheol1986,FoersterPhysFluids1994,SchaferDippelWolf1996}\\
Friction coefficient\footnotemark[1] $\mu$ & 0.7\\
Restitution coefficient\footnotemark[1] $e$ & 0.87\\
Fill fraction $f$ & 20\%-70\%\\
Collision time $\Delta t$ (s) & $10^{-4}$~\cite{Ristow2000,CampbellJFM2002,SilbertPRL2007}\\
Integration time step (s) & $\Delta t/50 = 2\cdot10^{-6}$~\cite{SchaferDippelWolf1996}\\
Stiffness coefficient $k_n$ (N\,m$^{-1}$) & $7.32\cdot10^{4}$~\cite{SchaferDippelWolf1996}\\
Damping coefficient $\gamma_n$ (kg\,s$^{-1}$) & 0.206\\
\end{tabular}
\end{ruledtabular}
\footnotetext[1]{For particle-particle and particle-wall interactions.}
\end{table}

\begin{figure*}[t!]
\begin{center}
\includegraphics[width=\textwidth]{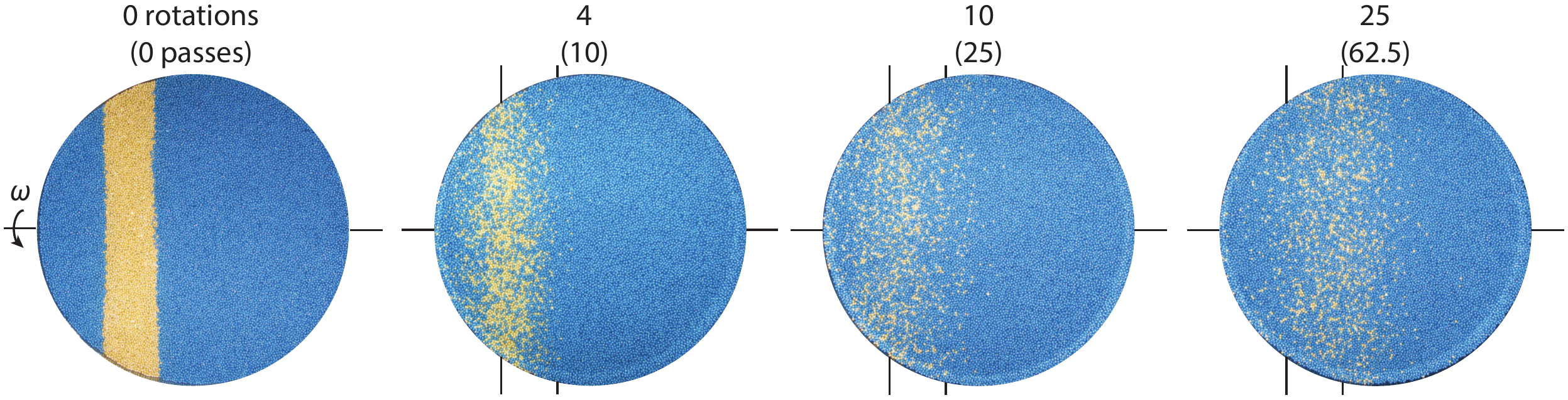}
\caption{(Color online) Top view of a colored band in a spherical tumbler for increasing rotation from experiment (initial band position denoted by vertical lines). Rotation number and approximate number of passes through the flowing layer (in parenthesis) are indicated above each image. $f = 30\%$, $d = 1$\,mm, $D = 14$\,cm, and $\omega= 5$\,rpm.}
\label{fighemisphere}
\end{center}
\end{figure*}

Cross sections of the bed in Fig.~\ref{figcrosssection}(b) after three rotations for $f=30\%$ reveal axial drift of particles, although collisional diffusion has somewhat blurred the band boundaries. The deformation of the band at and beneath the free surface indicates the direction and degree of axial drift. The band shifts toward the pole at the top surface and at the bottom wall of the tumbler, while it shifts toward the equator in the mid-layer. White dashed lines for the cross-section 1\,cm behind the meridional cut highlight the band deformation. Particles in the band pass through the flowing layer about eight times for $f=30\%$ after three tumbler rotations. Thus, axial drift is relatively small compared to the streamwise particle displacements. However, it is large enough to dominate over axial mixing due to collisional diffusion.

The colored band experiment is replicated using DEM simulations (see Appendix~\ref{simmethod} for details and Table~\ref{simtable} for parameters). Figure~\ref{figcrosssection}(c) demonstrates that tracer band deformation in the simulation is similar to that in the experiment: particles drift toward the poles at the free surface and toward the equator lower in the flowing layer. (see Supplemental Material~\footnote{\label{video} See Supplemental Material \href{http://link.aps.org/supplemental/10.1103/PhysRevE.88.012208} for a video of the band deformation over time.}) Particles near the surface and tumbler wall in the simulation drift further toward the pole than in the experiment, possibly due to differences in particle properties (i.e., $d$~\cite{PengfeiPRLBandInv2009}, $e$, and $\mu$) between the experiment and the simulation. Additional experiments at $f=40\%$ and $f=50\%$ and simulations at $f=20\%-70\%$ (see Fig.~\ref{figfillfrac}) exhibit similar axial drift.  Qualitatively similar drift for fill fractions equal to, less than, and greater than $50\%$ rules out potential mechanisms for drift related to axial variation of cycle time, as shown in Fig.~\ref{figtime}.

Further insight into axial drift was obtained from images of the free surface distribution of the light colored particles initially in the tracking band (Fig.~\ref{fighemisphere}).  After four rotations, the band is closer to the nearest pole and becomes wider and less distinct. After 25 rotations, beads from the band are closer to the equator. Collisional mixing of light and dark colored particles is substantial, but light colored particles remain almost exclusively in the hemisphere of the tumbler in which they started (left). To quantify this surface motion, the streamwise-averaged image intensity of the light colored beads was fit to a Gaussian. As determined by the peak of the Gaussian, centers of the light colored band are $z=[-3.05,-3.86,-3.00,-2.35]$\,cm for 0, 4, 10 and 25 rotations (respectively). Assuming that mean surface velocities are constant, the non-monotonic motion of the band center suggests that, in addition to axial flow, particles also move vertically through the layer, descending at the poles and ascending near the equator (see Sec.~\ref{meantraj}).

\begin{figure}[h!]
\begin{center}
\includegraphics[width=0.75\linewidth]{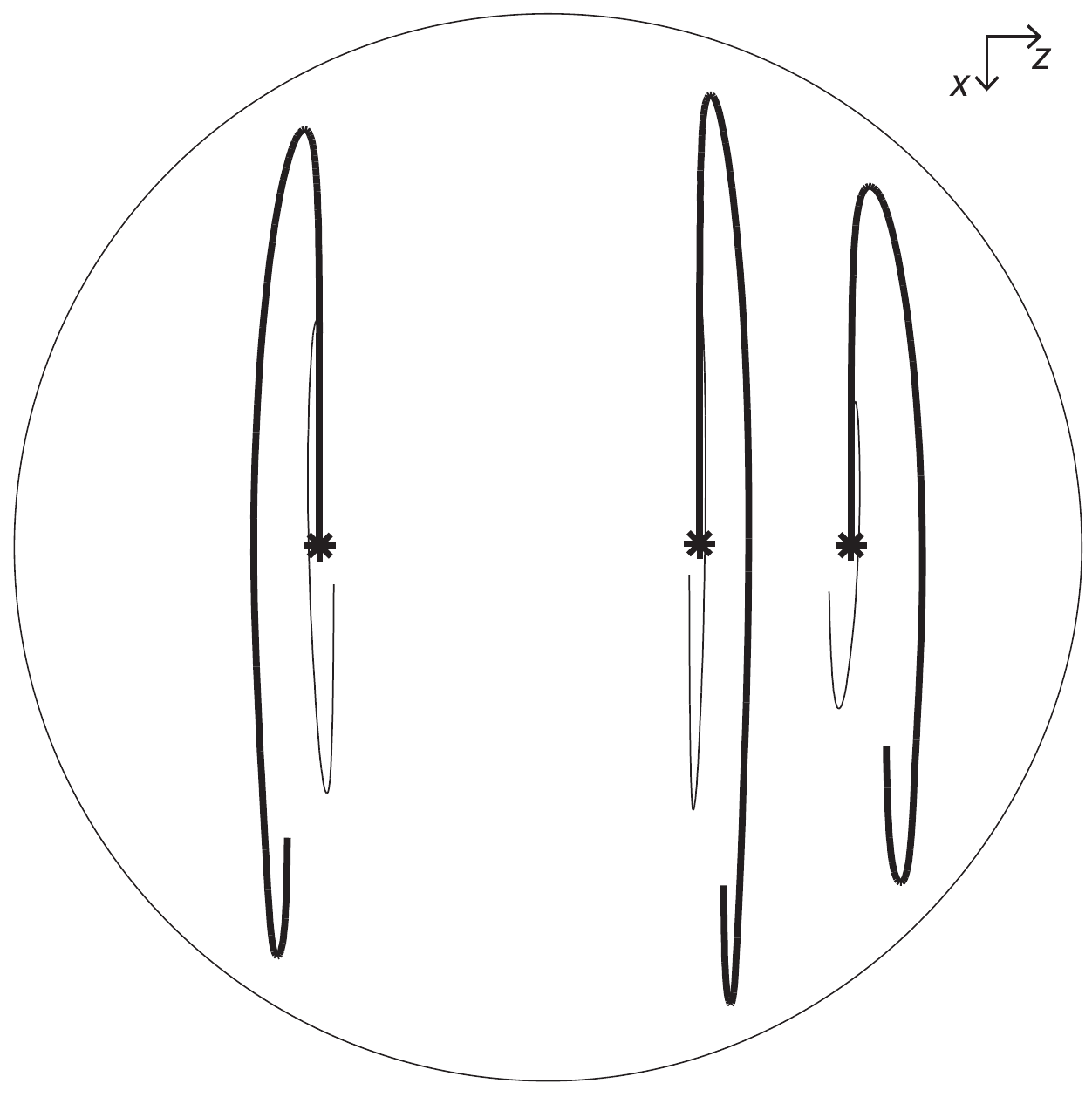}
\caption{Mean particle trajectories from simulation at two different initial depths viewed in the $xz$-plane exhibit net axial displacement per pass through the flowing layer. Bold (light) trajectories start (starting location denoted by asterisks) 0.2\,cm (1.8\,cm) from the tumbler wall in the fixed bed and traverse the flowing layer near the top (bottom) of the layer. $d=2$\,mm, $D=14$\,cm, $f=30\%$, and $\omega=30$\,rpm.}
\label{figtrajdepth}
\end{center}
\end{figure}

\begin{figure*}[!t]
\begin{center}
\includegraphics[width=\textwidth]{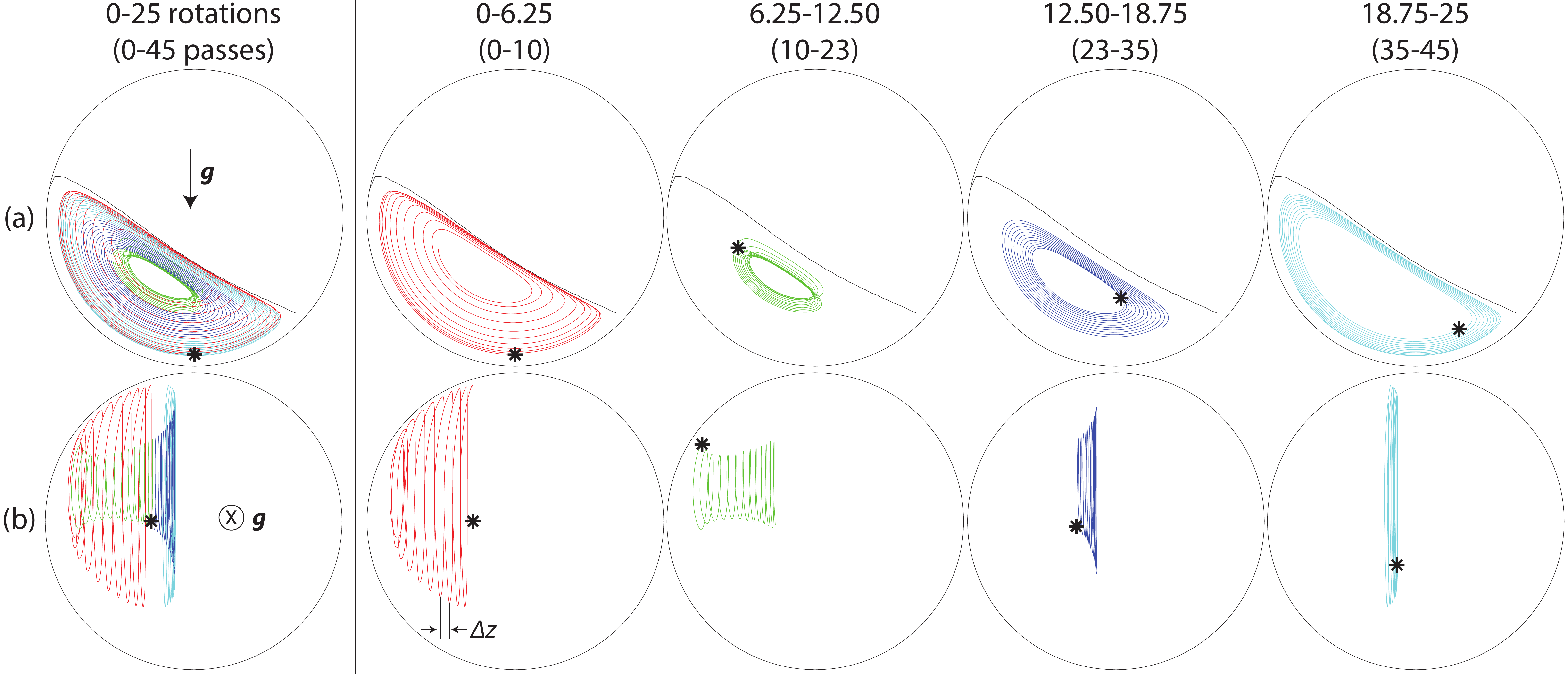}
\caption{(Color online) A representative mean particle trajectory for 25 tumbler rotations (45 flowing layer passes) from a spherical tumbler simulation with $D=14$\,cm viewed from (a) the side ($xy$-plane) and (b) the top (parallel to $\vec{g}$); tumbler rotations (passes) are noted above each column. The trajectory is shown in its entirety (first column) and in 6.25 tumbler rotation segments. The free surface is indicated by the black curve in (a), and the start of each trajectory section is indicated by an asterisk (*). Straight portions of the trajectory in (b) correspond to solid body rotation, while curved portions correspond to flow. The full trajectory starts 6.8\,mm from the tumbler wall at $z/R=-2/7$. It advances toward the pole through the top of the flowing layer and returns back to the equator through the bottom of the flowing layer.  $f=30\%$, $d=2$\,mm, and $\omega=30$\,rpm.}
\label{figtorus}
\end{center}
\end{figure*}

\section{Drift Kinematics}

\subsection{Mean trajectories}
\label{meantraj}
To better understand the axial drift, we examine average particle trajectories constructed by integrating the mean velocity field from simulation using a second order Runge-Kutta method. Figure~\ref{figtrajdepth} shows the effects of initial particle depth on pairs of trajectories at three axial locations viewed in the $xz$ plane (looking down onto the free surface). The two trajectories in each pair start (indicated by a filled circle) in the fixed bed and the $yz$ plane 0.2\,cm and 1.8\,cm away from the tumbler wall and enter the flowing layer near the free surface (bold curve) and near the bottom of the flowing layer (light curve), respectively. Near surface trajectories enter the flowing layer further upstream and exit it further downstream compared to deeper trajectories. More importantly, near surface trajectories move further toward the pole in the upstream half of the flowing layer than away from it in the downstream half of the flowing layer resulting in a net poleward motion as previously reported for segregating particles~\cite{PengfeiPRLBandInv2009}. In contrast, deeper trajectories exhibit the opposite behavior and move closer to the equator each pass. Axial displacement is greatest at the upstream and downstream ends of the flowing layer, while the trajectory in the middle portion of the flowing layer is nearly straight.

\begin{figure*}[!t]
\begin{center}
\includegraphics[width=\textwidth]{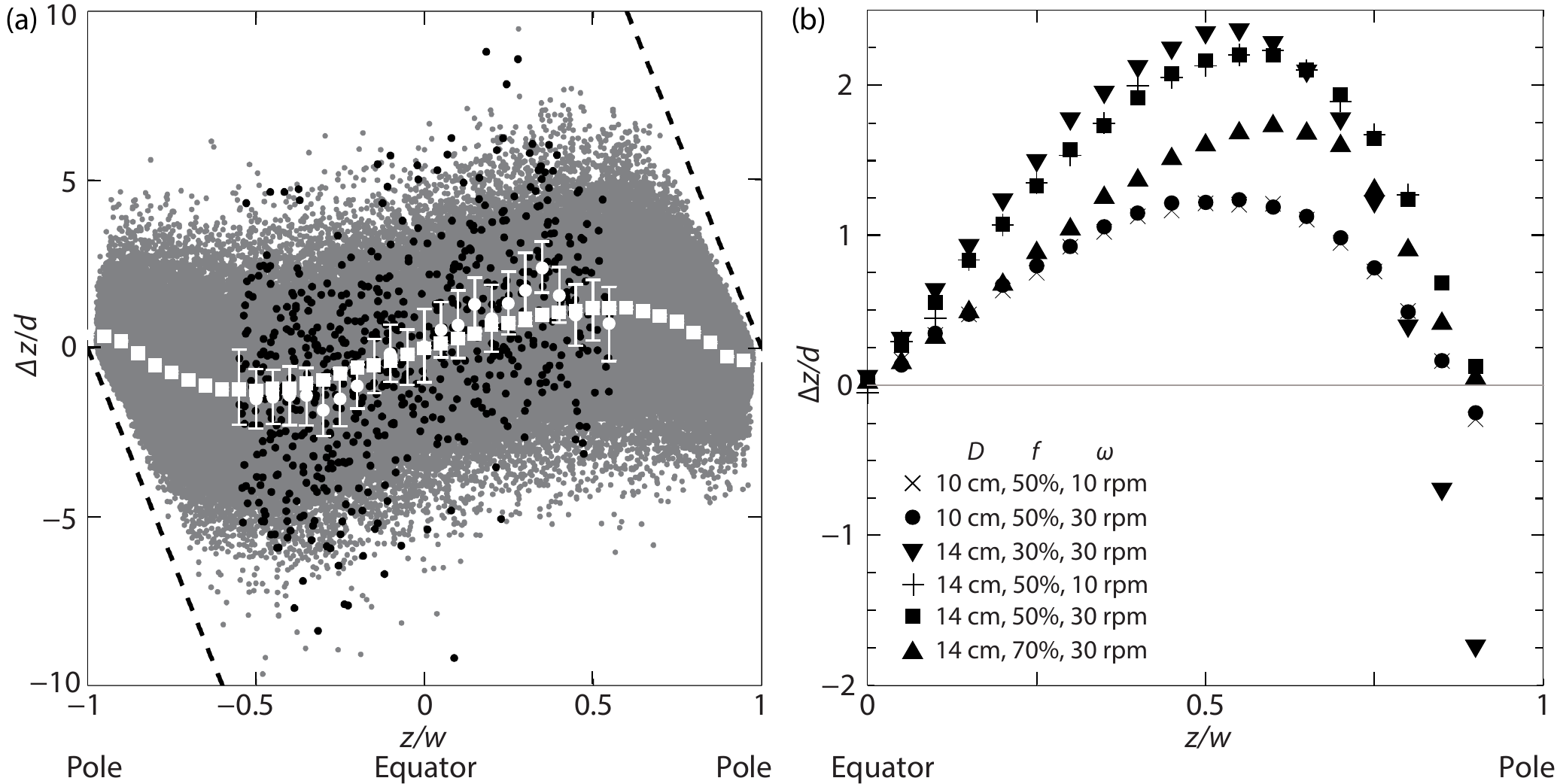}
\caption{Non-dimensional axial surface drift per pass $\Delta z/d$ vs.\ normalized axial position $z/w$ in a spherical tumbler with $d=2$-mm particles. (a) $\Delta z/d$ from experiment (black points) and simulation (gray points) are similar. Mean axial drift with 95\% confidence interval error bars for experiment (white circles) and for simulation (white squares) match within uncertainty (error bars for simulation are smaller than the symbols). Axial distance to the poles (dashed lines) bounds the axial displacement. $f=50\%$ and $\omega=10$\,rpm. (b) Mean axial surface drift in spherical tumbler simulations increases with $D$. $w$ is the half width of the free surface.}
\label{driftparticle}
\end{center}
\end{figure*}

To understand the global motion resulting from the axial drift per pass $\Delta z$, Fig.~\ref{figtorus} presents a nearly complete mean particle trajectory for 25 tumbler rotations and the same trajectory in shorter (6.25 rotation) increments. $\Delta z$ manifests as the spacing between the trajectory loops. In the top row, the view along the axis of rotation ($xy$ plane) shows the dynamic angle of repose of the particles as well as the depth of the particle trajectory. In the bottom row, the view is down the gravity vector with flow from top to bottom, so the trajectory is asymmetric with respect to the horizontal axis of rotation due to the $23^\mathrm{o}$ repose angle of the tumbling particles.

From 0 to 6.25 tumbler rotations, the trajectory drifts toward the pole with each cycle through the flowing layer and fixed bed, and the axial displacement with each loop increases slightly approaching the pole. In the top view, trajectories are straight during solid body rotation in the fixed bed and curved in the flowing layer where the axial position changes. From 6.25 to 12.50 tumbler rotations, the trajectory is deep in the flowing layer and migrates back toward the equator, as is evident from the side view. From 12.50 to 18.75 tumbler rotations, the trajectory continues to move toward the equator, but with decreasing axial displacement, as it simultaneously approaches the free surface. From 18.75 to 25 tumbler rotations, the trajectory drifts slowly back toward the pole through the top of the flowing layer. For the entire trajectory, the axial displacement per pass is small near the equator and pole and larger in between. The axial drift behavior in Fig.~\ref{figtorus} is consistent with that in Fig.~\ref{figtrajdepth}, where poleward drift occurs when the particle is near the surface of the flowing layer, and drift toward the equator occurs when the particle is deep in the flowing layer. The trajectory circulates through the domain of the tumbler. The poleward axial flux of particles near the surface of the flowing layer is balanced by the axial flux of particles toward the equator near the bottom of the flowing layer.

\subsection{Surface Drift}
\label{surfdrift}
The trajectory results reported in the previous section are derived from average velocity fields.  Here we consider the axial drift per pass, $\Delta z,$ of individual particles at the surface in experiments and simulations to better understand the stochastic nature of the drift process. $\Delta z$ is defined as the change in axial position of a surface particle in the fixed bed after a single pass through the flowing layer.  In simulation, particles in the fixed bed whose centers remain within  1.5\,mm (0.75$d$) of the tumbler wall after passing through the flowing layer are considered surface particles. Figure~\ref{driftparticle}(a) shows that $\Delta z/d$ measured from simulation (gray points) is stochastic with a nearly constant width of about 10 except near the poles where it is reduced due to the finite size of the tumbler which restricts the magnitude of poleward displacement (dashed lines). Axial displacements near the surface display a bias that depends on axial position. Poleward displacements outnumber equator directed displacements except near the pole ($z/w>0.9$) where particles are more likely to drift toward the equator. To quantify the mean drift, individual displacements are grouped into bins of width $1.25 d$ and the mean of each bin are computed (white squares in Fig.~\ref{driftparticle}(a)). Error bars representing the 95\% confidence interval are smaller than the symbols due to the large number of samples in each bin. The individual displacements are well fit by a cubic polynomial~\footnote{The weighted cubic polynomial was accepted as an appropriate fit for the mean axial drift using the F-test~\cite{Kutner2005} for lack of fit at the 1\% significance level. The individual displacements were weighted using sampling weights~\cite{Lohr2009} $N/n$, where $N$ is the total number of data points and $n$ is the number of individual displacements in the bin to which the data point belongs. The sampling weight allows every axial position to be considered equally regardless of axial sampling variations.}. The mean axial drift per cycle is zero at the equator (presumably due to symmetry of the tumbler about this plane), increases to a maximum of about $1.2d$ at $z/w \approx 0.6$, and then decreases to $-0.2d$ reflecting the eventual reversal of the drift for particles near the pole (due to the geometric limits on poleward displacement as shown by the dashed lines in Fig.~\ref{driftparticle}(a)).

Quantifying axial drift in experiment is more challenging because all particles cannot be easily tracked. Instead, a single tracer particle identical except for color was tracked at the clear wall of the tumbler in the fixed bed. A camera, mounted below the spherical tumbler, imaged fixed bed particles in contact with the wall at 30 fps. Since a single tracer appears at the wall infrequently, $O(10 000)$ tumbler rotations were required to obtain a minimum of 20 sets of observations (pairs of tracer sightings within 6 passes) per bin of length $1.25 d$. In comparison, ~5000 observations per bin were obtained by tracking all particles at the wall in simulation. Moreover, in experiments, tracer displacements could only be obtained for $-0.5 < z/w < 0.5$ due to the camera's inability to image the steeply curved tumbler wall near the pole. However, within this restricted region, the distribution of individual axial displacements from experiment (black points in Fig.~\ref{driftparticle}(a)) is similar to that from simulation (gray points) as is the mean axial drift (white circles calculated as in the simulation) whose error bounds overlap the mean axial drift from simulation (white squares). Since results from experiments and simulations are similar, simulations are again used to further probe axial drift at the surface.

The effects of varying tumbler diameter, speed, and fill fraction on the mean axial surface drift from simulations are shown in Fig.~\ref{driftparticle}(b) for the right hemisphere of the tumbler. All simulations demonstrate the same qualitative behavior: drift is 0 at the equator and increases approximately linearly to a maximum around $z/w = 0.5$ (except for $f=70\%$, which occurs slightly further from the equator). For $f=50\%$, maximum axial drift of the $D=14$\,cm simulation is 1.8 times that in the $D=10$\,cm simulation. Varying $\omega$ from 10\,rpm to 30\,rpm for both values of $D$ has no impact on the axial drift per pass. For $D=14$\,cm, the maximum axial drift is highest for $f=30\%$, just slightly greater than that for $f=50\%$ but significantly greater than $f=70\%$ despite having the same free surface geometry as $f=30\%$. The reduced drift at $f=70\%$ may be influenced by differences such as the greater bed depth near the pole and the existence of an unmixed core, which never enters the flowing layer~\cite{JainPhysFluids2002,BonamyPhysFluids2002,SociePRE2005}. In addition, the axial drift depends on particle properties. Increasing the restitution coefficient in simulations from 0.5 to 0.8 decreases the maximum axial drift by 30\%, while increasing the friction coefficient from 0.5 to 0.9 decreases the maximum axial drift by 16\%.

\begin{figure}[!h]
\begin{center}
\includegraphics[width=1\linewidth]{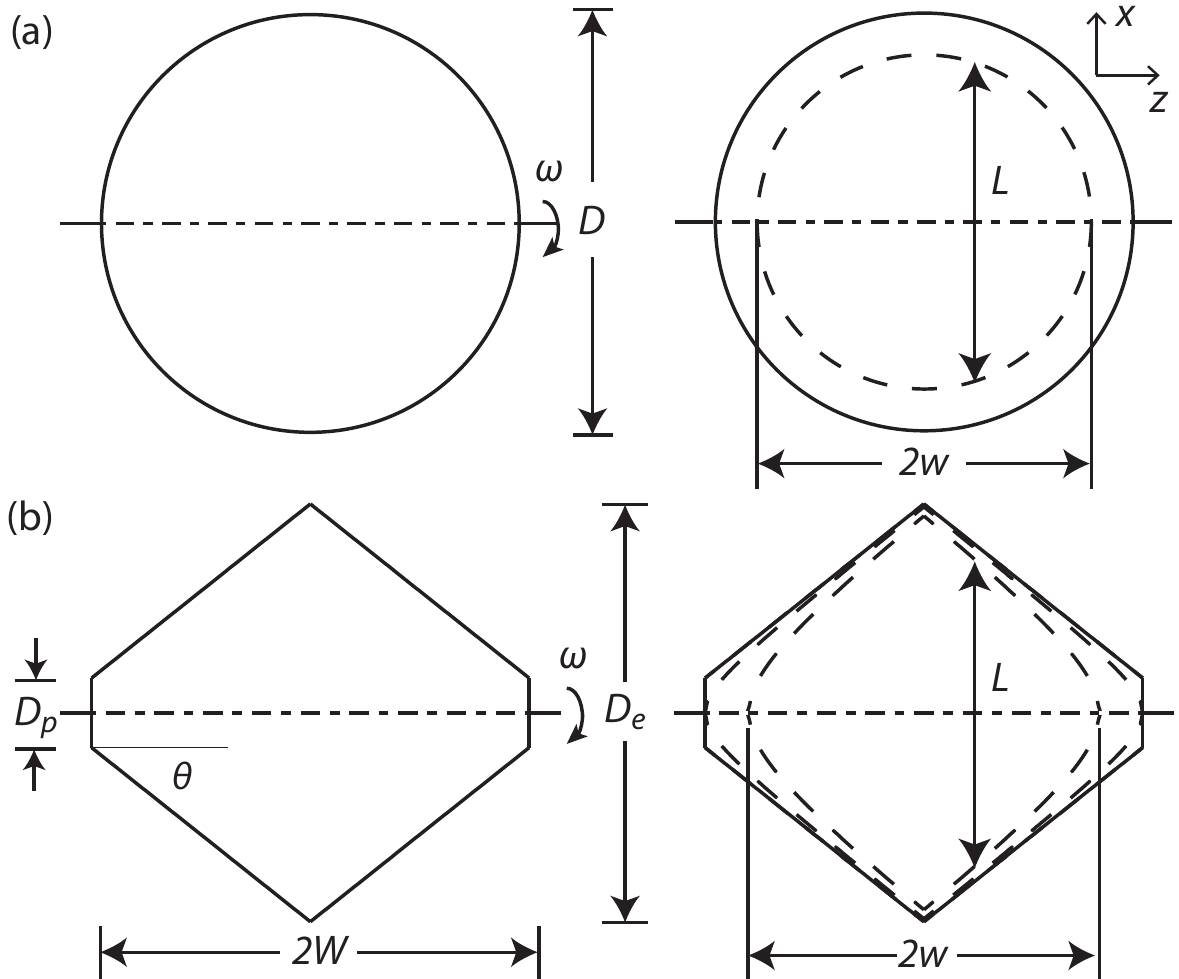}
\caption{Top views of (a)~spherical tumbler of diameter, $D$, and (b)~truncated double cone tumbler of width, $2W,$ equator diameter, $D_e$, and pole diameter, $D_p$. Rotation is about the $z$-axis, which is also the axis of symmetry for the double cone tumbler. (Left column) Bed cross sections at fill fraction $f=50\%$. (Right column) Sketches of the flowing layer surface (dashed curves) in the $xz$-plane for $f\neq50\%$ with flowing layer width, $2w$, along the axis of rotation.  The flowing layer length, $L$, varies axially for both spherical and conical tumblers. $2W$ is the axial width of the tumbler while $2w$ is the axial distance that the free surface extends in the tumbler, which may be less than the width of the tumbler depending of the fill level.}
\label{figflowlength}
\end{center}
\end{figure}

\begin{figure*}[!t]
\begin{center}
\includegraphics[width=\textwidth]{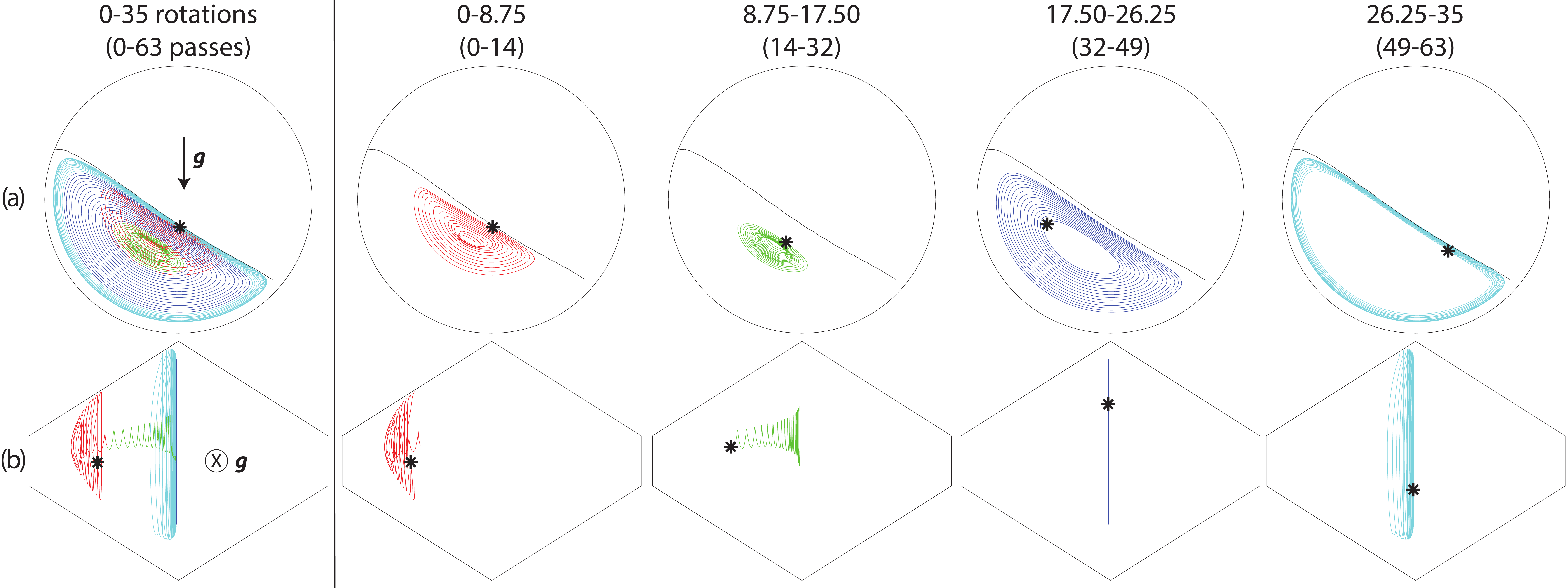}
\caption{(Color Online) Mean particle trajectory for a double cone tumbler simulation: (a) side view ($xy$ plane) and (b) top view (parallel to $\vec{g}$). Trajectory is shown for 35 rotations (63 flowing layer passes) in its entirety (first column) and in 8.75 tumbler rotation segments with $2W=15$\,cm, $D_{p}=2.5$\,cm, $D_{e}=12$\,cm, and $\theta=26.6^{\mathrm{o}}$. Tumbler rotations are given above each column with flowing layer passes in parentheses. Similar to the spherical tumbler (Fig.~\ref{figtorus}), straight portions of the trajectory in (b) correspond to solid body rotation while curved portions correspond to flow~\cite{Note4}. The free surface is indicated by the black curves in (a). The trajectory starts 0.6~mm below the surface and directly below the axis of rotation at $z/W=0.53$ (asterisk), advances towards the pole through the top of the flowing layer, and returns to the equator through the bottom of the flowing layer. $f=30\%$, $d=2$\,mm, and $\omega=30$\,rpm.}
\label{figtoruscone}
\end{center}
\end{figure*}

\section{Tumbler geometry}

While the experimental and simulation results for spherical tumbler flow reveal an axial drift that depends on axial position, they leave unresolved the mechanism of axial drift. Previous results \cite{SantomasoCES2004,PengfeiPRESubsurface2008} indicate that axial drift is negligible in cylindrical tumblers (except near the endwalls where wall friction drives the flow), suggesting that axial drift depends on axial variation of the flowing layer length, $L,$ or wall angle, $\theta,$ measured relative to the rotation axis in the plane of the free surface. To explore the effects of wall geometry on axial drift, simulations were done in double cone tumblers because in this geometry $L$ varies with axial position while $\theta$ remains constant (Fig.~\ref{figflowlength}).

We first note that the perimeter of the free surface in the $xz$-plane of spherical and double cone tumblers (Fig.~\ref{figflowlength}) depends on $f$. For $f=50\%,$ the free surface perimeter (solid curves) is characterized by the tumbler diameter ($D$ for spheres and $D_e$ and $D_p$ for double cones) and by the tumbler width ($D$ for spheres and $2W$ for double cones). However, for $f\neq50\%$, the free surface perimeter (dashed curves in the right column of Fig.~\ref{figflowlength}) does not necessarily span the width of the tumbler, resulting in an effective free surface width $2w.$ Regardless of $f$, the free surface in a spherical tumbler retains the same shape (circular) unlike the double cone tumbler whose free surface is bounded by two hyperbolas that depend on $f$~\footnote{Flow in double cones rotated about the $x$-axis (in Fig.~\ref{figflowlength}) has been explored~\cite{MoakherPowderTech2000,BronePowderTech2000,AlexanderPoF2001}. However, in this orientation, the flowing layer wall boundary changes shape from a circle to a double cone cross section during tumbling---an effect outside the scope of this study. We consider the double cone rotated about only the $z$-axis (Fig.~\ref{figflowlength}(b)) where the flowing layer retains the same shape during rotation.}. Unless otherwise noted, for the double cone geometries considered, the fill level ($f=30\%$) was chosen such that the free surface extended just to the endwalls, corresponding to the outermost dashed curves for the $2w=2W$ case in Fig.~\ref{figflowlength}(b). Additionally, the double cone tumbler simulation parameters are identical to those used for the spherical tumbler simulation (Table~\ref{simtable})~\footnote{For ease of simulation, the double cone walls were defined by a monolayer of fixed particles identical to the tumbler particles rather than a smooth surface as used for the spherical tumbler simulations.}.

In Fig.~\ref{figtoruscone}, particle trajectories calculated by integration of the mean velocity field in simulation show that axial drift in a double cone tumbler ($D_{e}=12$\,cm, $D_{p}=2.5$\,cm, $2W=15$\,cm, and $\theta = 26.6^{\mathrm{o}}$) is qualitatively similar to that in a spherical tumbler (Fig.~\ref{figtorus}). A trajectory starting in the top of the flowing layer advances toward the pole and returns to the equator lower in the flowing layer~\footnote{\label{ftnote}The apparent axial drift in the fixed bed (during upward motion in the figure) for rotations 8.75–17.50 may be a consequence of
two effects, neither of which represent actual flow in the fixed bed. First, the trajectory may be located at the very bottom of the flowing layer where the apparent reverse flow comes about because the total velocity is the sum of the very slow positive streamwise velocity of the flowing layer superimposed on the slightly faster negative velocity due to the overall rotation of the tumbler system. Second, it is difficult to spatially resolve the velocity field very near the boundary between the flowing layer and the fixed bed since discrete bins used for averaging the velocity field may locally overlap both the flowing layer and the fixed bed}. Near the equator, it slowly returns to the top of the flowing layer and repeats the process. The persistence of drift in the double cone tumbler indicates that axially varying wall curvature is not required for axial drift to occur.

\begin{figure}
\begin{center}
\includegraphics[width=1\linewidth]{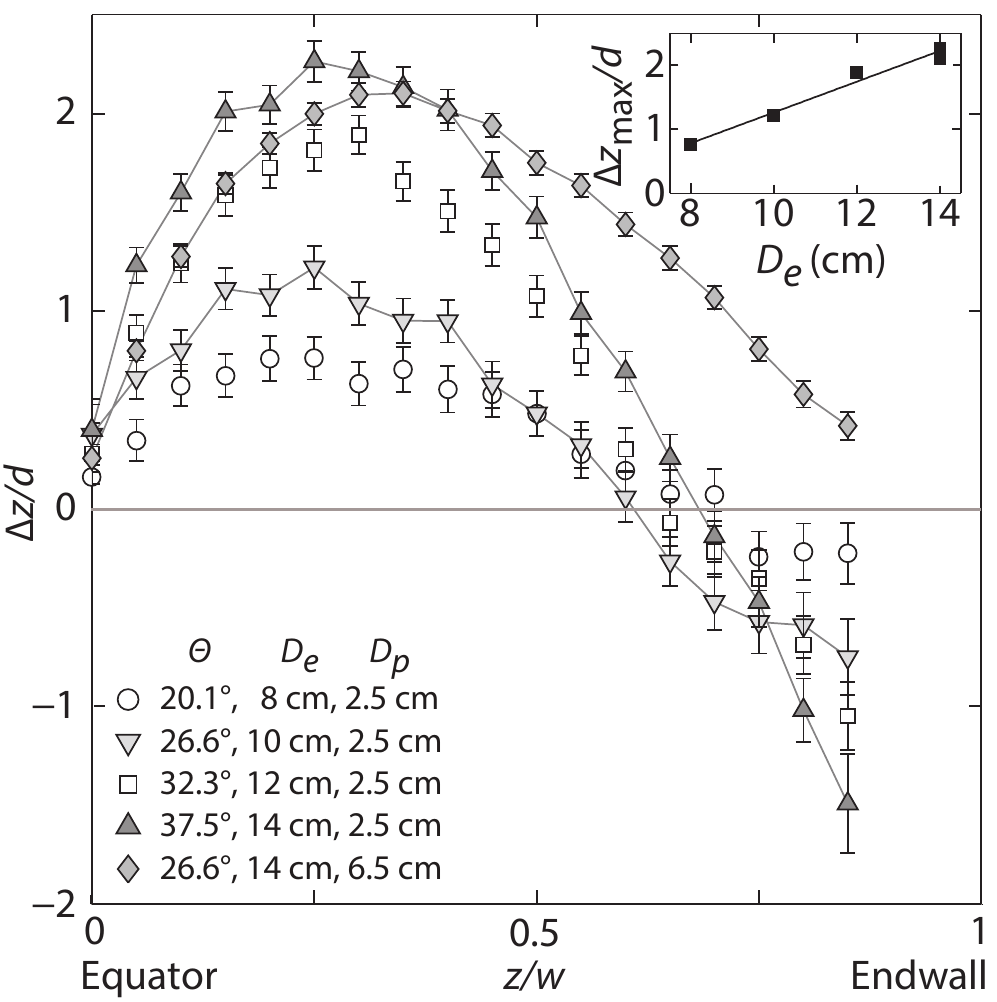}
\caption{Non-dimensional mean axial surface drift $\Delta z/d$ of $d=2$\,mm particles in simulation vs.~normalized axial position $z/w$ measured at the double conical tumbler wall by particle tracking as in Fig.~\ref{driftparticle} ($f=30\%$, $d=2$\,mm, $\omega=30$\,rpm). Lines connecting data points emphasize three simulations that are compared directly for wall slope and equator diameter effects. Inset: Maximum dimensionless axial displacement increases linearly with $D_e$. All simulations have $w=W=7.5$\,cm, since the free surface extends from endwall to endwall.}
\label{axialdisp}
\end{center}
\end{figure}

Comparing Fig.~\ref{figtoruscone} for a double cone tumbler with the analogous results for the spherical tumbler in Fig.~\ref{figtorus} raises the question of whether $\theta$ or $L$ determines the magnitude of axial drift. To address this question, we simulated axial drift in double cone tumblers with different $\theta$. This was accomplished by holding $D_p$ and $W$ constant and selecting equator diameters, $D_{e},$ of 8, 10, 12, and 14\,cm, which correspond to wall slope angles, $\theta$, of $20.1^{\mathrm{o}}$, $26.6^{\mathrm{o}}$, $32.3^{\mathrm{o}}$, and $37.5^{\mathrm{o}}$, respectively. An additional simulation with $D_e = 14$\,cm, $D_p = 6.5$\,cm, and $\theta = 26.6^{\mathrm{o}}$ was performed to further investigate the role of $D_e$ and $\theta$ on the axial drift. For the simulations with ($D_e = 8$\,cm, $D_p = 2.5$\,cm) and ($D_e = 14$\,cm, $D_p = 6.5$\,cm), particles filled the tumbler above where the endwall meets the conical section.

Axial drift in the double cone tumbler simulations was measured with the same particle tracking procedure used for the spherical tumbler simulations (Fig.~\ref{driftparticle}). $\Delta z/d$ is plotted in Fig.~\ref{axialdisp} as a function of $z/w$ for all the double cone simulations. Like the spherical tumbler simulations, $\Delta z$ is 0 at the equator and peaks away from the equator. However, the location of the peak, $z_{\mathrm{max}}$, occurs closer to the equator in the double cones ($z/w\approx0.25$) than in the sphere ($z/R\approx0.6$). The maximum of $\Delta z$, $\Delta z_{\mathrm{max}}$, increases linearly with $D_e$ (Fig.~\ref{axialdisp} inset) independent of $\theta$. The linear scaling of $\Delta z_{\mathrm{max}}$ with $D_e$ is expected to fail for small $D_e$ as the bed of particles becomes shallow. In addition, $\Delta z_{\mathrm{max}}$ for the double cones with $D_e = 10$\,cm and $D_e = 14$\,cm (Fig.~\ref{axialdisp}) is nearly the same as for the spheres (Fig.~\ref{driftparticle}) with corresponding $D$, suggesting that equator diameter controls $\Delta z_{\mathrm{max}}$ in both geometries. Near the pole, the axial drift is more negative than in the spherical tumbler (except for the $D_p=6.5$\,cm case), likely due to the relative shallowness of the particle bed in this region compared to that of the spherical tumbler.

To further elucidate the role of $\theta$ in the drift, we focus on the simulations with ($D_e = 10$\,cm, $D_p = 2.5$\,cm), ($D_e = 14$\,cm, $D_p = 2.5$\,cm), and ($D_e = 14$\,cm, $D_p = 6.5$\,cm) (denoted by filled symbols in Fig.~\ref{axialdisp}). The wall angle for ($D_e = 14$\,cm, $D_p = 6.5$\,cm) and ($D_e = 10$\,cm, $D_p=2.5$) is the same, and both tumblers exhibit nearly the same rate of decrease of axial drift with axial position from the maximum to the endwall indicating that wall slope determines this behavior. As mentioned earlier, the maximum axial drift is controlled by $D_e$. Thus, the maximum of axial drift depends on the length of the flowing layer, while the rate of change of $\Delta z$ with axial position from this maximum toward the pole depends on the wall slope. Moving from the maximum to the equator, $\Delta z$ decreases continuously to zero. $\Delta z$ is zero at the equator due to the reflection symmetry of the tumbler about the vertical midplane.

\begin{figure}
\begin{center}
\includegraphics[width=1\linewidth]{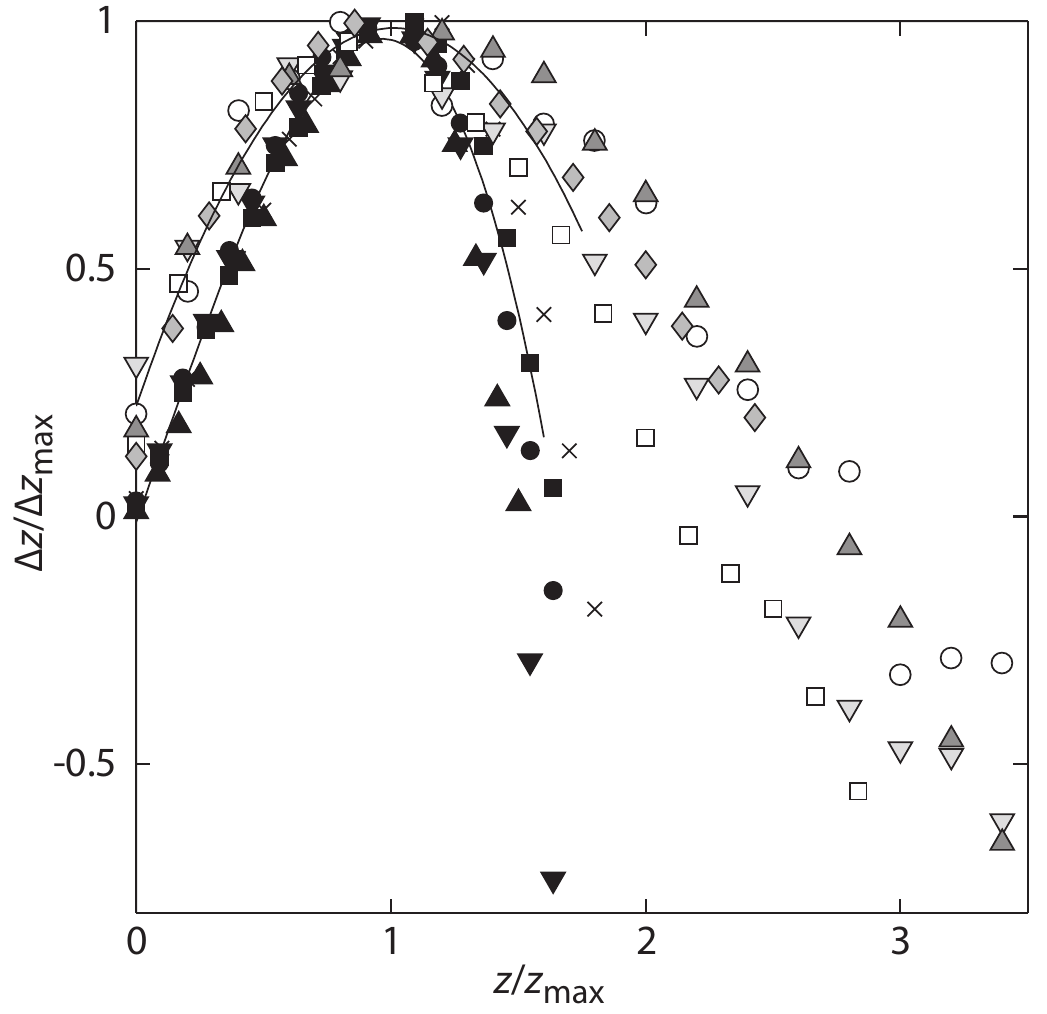}
\caption{Scaling of axial surface drift in double cone and spherical tumbler simulations (from data in Figs.~\ref{driftparticle}(b)~and~\ref{axialdisp}) using normalized axial surface drift, $\Delta z/\Delta z_{\mathrm{max}}$, and normalized axial position, $z/z_\mathrm{max}$. A parabolic fit for double cone data and a cubic fit for sphere data are shown by solid curves. Plot symbols are as in Fig.~\ref{driftparticle}(b) and Fig.~\ref{axialdisp}, where black filled symbols and $\times$'s are from spherical tumbler simulations and open and gray filled symbols are from double cone tumbler simulations.}
\label{axialdisp-collapse}
\end{center}
\end{figure}

In Fig.~\ref{axialdisp-collapse}, which compares axial drift in spherical and double cone tumblers, axial drift is normalized by $\Delta z_{\mathrm{max}},$ and axial position is normalized by the axial location of the maximum, $z_\mathrm{max}$. This scaling reveals a separate master curve for the axial drift in each tumbler geometry. While the two curves are similar, the different functional dependence of flowing layer length on axial position results in differences between the two curves. Attempts were made to use a physically motivated scaling for the drift but these scalings did not collapse the data well.

\section{Conclusion}

Spherical and double cone tumblers rotating about a horizontal axis exhibit weak three dimensional flow which takes the form of a slow axial drift. The axial drift is 1-2 particle diameters, corresponding to 1\%-3\% of the tumbler diameter, with each pass through the flowing layer. The drift is caused by asymmetries in the curvature of mean particle trajectories in the flowing layer and results in poleward drift near the free surface and equator directed drift deeper in the flowing layer. Drift is caused by the axial slope of the bounding walls, while the length of the flowing layer determines its magnitude. Axial drift is largely independent of wall friction as both smooth and rough walls yield similar results.

This research was motivated by the commonly used assumption in continuum models of granular flow that flow in any slice perpendicular to the rotation axis is independent of adjacent slices---in other words, the flow in a slice is two-dimensional. Implicitly, this assumption requires that the spanwise flow is negligible and any spanwise motion is primarily diffusive with zero mean.  The results here demonstrate that this assumption has limitations, since axial displacements of 1\%-3\% of the tumbler diameter occur with each pass through the flowing layer. It is unclear if axial drift would be present for particles and tumbler sizes not considered by this study as other forces, such as cohesion and electrostatic forces, may play a larger role. In some circumstances, axial drift might be used to an advantage. For example, it may be possible to offset the surface axial drift away from endwalls observed in cylindrical tumblers with flat endwalls \cite{SantomasoCES2004,PengfeiPRLBandInv2009} by using hemispherical (or other non-planar) endwalls which induce drift in the opposite direction.

While this work addresses aspects of axial drift in non-cylindrical tumblers, several details remain poorly understood. Specifically, the mechanism for axial drift in the flowing layer deserves further investigation. It is evident that, in a tumbler with axially sloped walls, poleward drift near the free surface and equator directed drift deeper in the flowing layer satisfy continuity. It is possible that the drift is initiated by particles in the free surface due to axial gradients in the streamwise velocity while particles deep in the flowing layer drift toward the equator to satisfy continuity. Particles do not cross the equatorial plane (except by collisional diffusion) in the spherical or double cone tumblers; this is clearly a consequence of the equator being a plane of symmetry with axial drift having opposite signs on opposite sides of the equator.  However,  it is not clear what would occur if the plane where the wall slope changes has a discontinuous and non-symmetric wall slope change, say for example, in the case of planes formed by joining a cylindrical tumbler with conical (or hemispherical) ends. It is also not apparent whether the drift occurs in the avalanching regime (at lower tumbler rotational speeds) and in the cataracting regime (curved free surface at higher tumbler rotational speeds). Additionally, it is not immediately clear how the axial variation in particle bed depth impacts the drift. Preliminary experiments at low fill fractions ($f<10\%$) have shown particle slip impacts particle trajectories in the flowing layer. These and other questions remain to be addressed.


\begin{acknowledgments}
This research was funded by NSF Grant CMMI-1000469. UDO thanks the M\'esocentre d'Aix-Marseille Universit\'e for providing computing facilities. Travel for RML was supported by the Carnot STAR Institute. Wendy Chan provided useful insight regarding the statistical analysis.
\end{acknowledgments}

\begin{table}[h!]
\caption{\label{timetable} Normalized solid body rotation time, $\tau/T$, and normalized flowing layer passage time, $\tau_{\mathrm{fl}}/T$, obtained by tracking a $d=3$\,mm tracer particle in a spherical tumbler ($f = 50\%$,~$D=14$\,cm,~$d=2$\,mm). $\sigma$ is the standard deviation.}
\begin{ruledtabular}
\begin{tabular}{lcccc}
Method & $\omega$ (rpm) & $\tau/T \pm \sigma$  & $\tau_{\mathrm{fl}}/T \pm \sigma$ & Passes \\
\hline
Experiment & 1 & $0.535\pm0.009$ & $0.066\pm0.007$ & 20\\
Experiment & 2 & $0.531\pm0.071$ & $0.065\pm0.015$ & 20\\
Experiment & 3 & $0.540\pm0.023$ & $0.072\pm0.014$ & 20\\
Simulation & 3 & $0.537\pm0.006$ & $0.066\pm0.007$ & 16\\
Simulation & 5 & $0.536\pm0.006$ & $0.072\pm0.007$ & 12\\
Simulation & 15 & $0.556\pm0.005$ & $0.101\pm0.005$ & 16\\
Simulation & 30 & $0.586\pm0.009$ & $0.129\pm0.006$ & 27\\
Simulation\footnotemark[1] & 30 & $0.576\pm0.009$ & $0.128\pm0.005$ & 69\\
\end{tabular}
\end{ruledtabular}
\footnotetext[1]{rough walls}
\end{table}

\appendix
\section{Residence Times}
\label{restime}

The negligible flowing layer passage time assumption made in Sec.~II was verified by experiments and simulations with $d=3$\,mm colored tracer bead in a bed of $d=2$\,mm beads. A larger tracer was used because it remained at the top of the flowing layer and at the tumbler wall when in solid body rotation, which allowed the time spent in solid body rotation to be measured. $f=50\%$ was used because $\tau/T$ should be independent of $z$ for this fill fraction. The relative flowing layer time, $\tau_{\mathrm{fl}}/T$, and the solid body rotation residence time, $\tau/T$, were measured for 20 tumbler rotations. Results in Table~\ref{timetable} indicate that $\tau/T$ is slightly longer than the predicted value of 0.5 for $f=50\%,$ probably due to a small degree of internal slip and slight rearrangement~\cite{KomatsuPRL2001,SociePRE2005,ArndtPRE2006} of particles in the fixed bed as the tumbler rotates. $\tau_{\mathrm{fl}}/T$ is about an order of magnitude less than $\tau/T$. It increases slightly with rotation rate, which is probably a result of the thicker flowing layer at these high flow rates. From simulations, the flowing layer at the equator was 1.3 times thicker at 30\,rpm than at 3\,rpm, resulting in a twofold increase in flowing layer passage time. Flowing layer passage times can also be estimated from experiments in a quasi-2D circular tumbler by Jain et al.~\cite{JainPhysFluids2002} and from the model of Christov et al.~\cite{ChristovSIAM}; both methods give values similar to those obtained in tracer experiments (Table~\ref{timetable}).
\\

\section{Simulation Method}
\label{simmethod}
For the DEM simulations, a standard linear-spring and viscous damper force model~\cite{CundallStrack1979,SchaferDippelWolf1996,Ristow2000,PengfeiPRESubsurface2008} was used to calculate the normal force between two contacting particles: $\mathbf{F}_{ij}^{n}=[k_{n}\delta-2\gamma_{n}m_{\mathrm{eff}}(\mathbf{V}_{ij}\cdot\hat{\mathbf{r}}_{ij})]\hat{\mathbf{r}}_{ij}$, where $\delta$ and $\mathbf{V}_{ij}$ are the particle overlap and the relative velocity ($\mathbf{V}_{i}-\mathbf{V}_{j}$) of contacting particles $i$ and $j$ respectively; $\hat{\mathbf{r}}_{ij}$ is the unit vector in the direction between particles $i$ and $j$; $m_{\mathrm{eff}}=m_{i}m_{j}/(m_{i}+m_{j})$ is the reduced mass of the two particles; $k_{n}=m_{\mathrm{eff}}[(\frac{\pi}{\Delta t})^2+\gamma_n^2]$ is the normal stiffness and $\gamma_{n}=\frac{\ln e}{\Delta t}$ is the normal damping, where $\Delta t$ is the collision time and $e$ is the restitution coefficient~\cite{Ristow2000,PengfeiPRESubsurface2008}. A standard tangential force model~\cite{CundallStrack1979,SchaferDippelWolf1996} with elasticity was implemented: $\mathbf{F}_{ij}^{t}=-\min(|\mu\mathbf{F}_{ij}^{n}|,|k_{s}\zeta|)\sign(\mathbf{V}_{ij}^{s})$, where $\mathbf{V}_{ij}^{s}$ is the relative tangential velocity of two particles~\cite{RapaportPRE2002}, $k_{s}$ is the tangential stiffness, and $\zeta(t)=\int_{t_{0}}^{t}\mathbf{V}_{ij}^{s}(t')\,{\rm {d}t'}$ is the net tangential displacement after contact is first established at time, $t=t_{0}$. The velocity-Verlet algorithm~\cite{Ristow2000,AllenTildesley2002} was used to update the position, orientation, and linear and angular velocity of each particle. Tumbler walls were modeled as both smooth surfaces (smooth walls) and as compositions of bonded particles (rough walls). Both wall conditions have infinite mass for calculation of the collision force between the tumbling particles and the wall. Rough walls were used exclusively for double cone tumbler simulations. Most spherical tumbler simulations used smooth walls, though a few used rough walls for comparison---both wall types produced similar results. To characterize the mean velocity field, the spherical computational domain was divided into cubical bins of width $1.3d$.  In most cases, this bin width adequately resolves the flowing layer, as the thickness of the flowing layer typically ranges from $5d-20d$~\cite{PignatelPRE2012}.  Local flow properties were obtained by averaging values for all particles in each bin every 100 time steps for a total of 15\,s of physical time.

\end{document}